\documentclass[fleqn,usenatbib]{mnras}

\usepackage{newtxtext,newtxmath}

\usepackage[T1]{fontenc}

\DeclareRobustCommand{\VAN}[3]{#2}
\let\VANthebibliography\thebibliography
\def\thebibliography{\DeclareRobustCommand{\VAN}[3]{##3}\VANthebibliography}

\usepackage{graphicx}	
\usepackage{amsmath}	

\title[4U~1210$-$64]{4U~1210$-$64: a new member of the rare intermediate-mass X-ray binary subclass}

\author[Monageng et al.]{
I.M. Monageng$^{1,2}$\thanks{E-mail: itu@saao.ac.za},
V.A. McBride$^{1,2,3}$,
J. Alfonso-Garzon$^{4}$,
L. J. Townsend$^{1,5}$,
  \newauthor 
J.B. Coley$^{6,7}$,
B. Montesinos$^{4}$,
R.H.D. Corbet$^{8,9}$,  
K. Pottschmidt$^{7,9}$,
\\
$^{1}$South African Astronomical Observatory, P.O Box 9, Observatory, 7935, Cape Town, South Africa\\
$^{2}$Department of Astronomy, University of Cape Town, Private Bag X3, Rondebosch 7701, South Africa\\
$^{3}$International Astronomical Union's Office of Astronomy for Development, c/o SAAO\\
$^{4}$Centro de
Astrobiologia (CAB), INTA-CSIC, Camino Bajo del Castillo s/n, E-28692
Villanueva de la Cañada, Madrid, Spain\\
$^{5}$Southern African Large Telescope, P.O Box 9, Observatory, 7935, Cape Town, South Africa\\
$^{6}$Department of Physics and Astronomy, Howard University, Washington, DC 20059, USA \\
$^{7}$CRESST/Mail Code 661, Astroparticle Physics Laboratory, NASA Goddard Space Flight Center, Greenbelt, MD 20771, USA \\
$^{8}$CRESST/Mail Code 662, X-ray Astrophysics Laboratory, NASA Goddard Space Flight Center, Greenbelt, MD 20771, USA \\
$^{9}$Center for Space Sciences and Technology, University of Maryland, Baltimore County, Baltimore, MD 21250, USA\\
}

\date{Accepted XXX. Received YYY; in original form ZZZ}

\pubyear{2015}

\begin{document}
\label{firstpage}
\pagerange{\pageref{firstpage}--\pageref{lastpage}}
\maketitle

\begin{abstract}
\hbox{4U 1210$-$64} is a peculiar X-ray binary that was first discovered in 1978 by the \textit{Uhuru} satellite. The analysis of the X-ray data revealed a 6.7-day orbital period and an additional long-term modulation that is manifested as low and high flux states. Based on the previous classification of the donor star from the analysis of its optical spectra, the system has been suggested to be a high-mass X-ray binary. We re-visit the optical classification where, based on the spectra from the Southern African Large Telescope (SALT), we conclude that the donor star is of spectral class A8 III-IV, making it a member of the rare intermediate-mass X-ray binaries. We perform radial velocity analysis using the SALT spectra where we consider circular and eccentric orbits. From the mass function derived and the mass constraints of the donor star, we demonstrate that a neutron star is favoured as the compact object in the binary system. We show, for the first time, the folded optical lightcurves, whose shape is interpreted to be due to a combination of ellipsoidal variations, irradiation of the donor star, and mutual eclipses of the star and accretion disk.
\end{abstract}




\section{Introduction}

X-ray binary stars are interacting stellar systems comprising a donor star and a compact object (either a neutron star or a black hole) that is accreting matter from its companion. Depending on the nature of the donor star, X-ray binaries are generally classified into low-mass X-ray binaries (for systems that have a late-type donor, with mass $\lesssim 1$~M$_\odot$) and high-mass X-ray binaries (for cases where the donor is an early-type with mass $\gtrsim 10$~M$_\odot$). Companions with masses in between these extremes, the intermediate-mass X-ray binaries, are rare \citep{Tauris2006}. It has been suggested that the intermediate-mass X-ray binary phase represents a stage just before the low-mass X-ray binary phase, where most of the mass from the donor star has not yet been lost/transferred \citep{1995JApA...16..255V,2002ApJ...565.1107P,2003ApJ...597.1036P}. Two of the most well-known intermediate-mass X-ray binary systems are Cyg~X-2 (neutron star and a companion of spectral-type A9; \citealt{1998ApJ...493L..39C}) and Her~X-1 (neutron star with an A7 spectral-type companion; \citealt{2014ApJ...793...79L}).\\

The interaction between the donor star and the compact object can result in an increase in X-ray emission due to the accretion of matter from the donor onto the compact object. Depending on the nature of the two objects and their orbits, the process of accretion can take place through different mechanisms. Wind and/or circumstellar disc accretion is the most dominant form of mass transfer in high-mass X-ray binaries, while in the case of low-mass X-ray binaries matter is transferred via Roche-lobe overflow. 

4U 1210$-$64, the subject of this paper, was first discovered by the \emph{Uhuru} satellite in 1978 \citep{Forman1978}, and has suffered a few identity crises since. It has been reported both as a cataclysmic variable \citep{Revnivtsev2007} and a high-mass X-ray binary \citep{Masetti2010} in the past. Using the lightcurve from the \textsl{Rossi X-ray Timing Explorer (RXTE)} All-sky Monitor (ASM) \citet{Corbet2008} found a modulation of $\sim$6.7\,days, which they interpreted as the orbital period. From the analysis of the long-term ASM and MAXI light curves, \citet{Coley2014} found an X-ray eclipse with an eclipse half-angle of 11$^{\circ}_{.}$2$\pm$0$^{\circ}_{.}$6, which they estimated the spectral type of the mass donor in the binary to be a B5 III or earlier. In addition to the orbital modulation, the X-ray behaviour of 4U 1210$-$64 shows a long-term periodicity on the order of 5083~days that is indicative of a variable accretion rate,  where it undergoes high and low flux states \citep{Coley2014,Nakajima2022}. The analysis of the X-ray data shows that the eclipse profile is not observed during the low flux state. The nature of the compact object in this system remains unclear, as no pulsations and cyclotron features have been detected. A detailed background of past X-ray observations is provided by \citet{Coley2014}. In this paper, we use the same ephemeris that is used in \citet{Coley2014}: $P = 6.7101 \pm 0.0005$ and $T_0 = \text{MJD}55553 \pm 0.002$. In this work, we present optical spectroscopy and photometry of the optical counterpart of \hbox{4U 1210$-$64}, where we re-visit the spectral type and the implications of the variability. 

This paper is structured as follows: we present the observations in Section~\ref{sec:observations}, and discuss the characterization of the donor star in Section~\ref{sec:donorstar}. In Section~\ref{sec:rv} we present the radial velocity analysis and the implications of the mass of the compact object. In Section~\ref{sec:photometry} we present the photometric variability. A discussion of the results is presented in Section~\ref{sec:discussion}, where we make inferences from the observational data, and summarise the paper in Section~\ref{sec:conclusion}.  

\section{Observations}
\label{sec:observations}
\subsection{Spectroscopy}
We observed the optical counterpart of 4U~1210$-$64 in two separate programmes in 2013 and 2022, where we collected 19 spectra between 2013 January and 2023 February using the Robert Stobie Spectrograph \citep{Burgh2003,Kobulnicky2003} on the Southern African Large Telescope \citep{Buckley2006}. The spectrograph was used in long-slit mode, with a slit width of $1.5^{\prime\prime} $, and employing grating PG0900 at an angle of $14.375^\circ$, with the camera station at $28.75^\circ$. This resulted in a wavelength range of 3884 -- 6987\,\AA~ and resolution of 6.6\, \AA \,at 4300\,\AA. Exposure times of 180\,s and 600\,s were used for data obtained in 2013 and 2022/3, respectively.\\
The PG3000 grating was used in one observation (2023-02-12) at a grating angle of $39.8725^\circ$ (camera station angle = $79.745^\circ$) which resulted in a wavelength range of 3875 -- 4610\,\AA\AA~ (resolution of 1.4\ \AA\, at 4300\,\AA). An exposure time of 1200\,s was used for the observation.

The observations were corrected for gain and amplifier cross-talk by the RSS pipeline \citep{Crawford2010}. Spectra were extracted using the standard longslit tools available in \textsc{iraf} \footnote{Image Reduction and Analysis Facility: iraf.noao.edu}, and extracted spectra had signal-to-noise ratios between 23 and 99 per pixel, dependent on the observing conditions. Further details on our observations are presented in Table\,\ref{tab:RV_table}.

\subsection{Photometry}
\label{subsec:phot}

We have compiled a range of photometric archival data.

\subsubsection{Gaia}
We found the Gaia counterpart of 4U 1210$-$64, which is identified as Gaia\,6053076566300433920. The source was classified as a Cepheid in Gaia data release 2 (Gaia DR2, \citealt{GaiaDR2}) providing an estimation of the period of 3.35560938\,d, which is half the orbital period of the system. We downloaded the Gaia G ($3300 - 10500$~\AA), BP ($3300 - 6800$~\AA), and RP ($6300 - 10500$~\AA) light curves from the catalogue I/345 through Vizier catalogue access tool \citep{Ochsenbein2000}.

\subsubsection{Las Cumbres Observatory}
We obtained photometric observations of 4U 1210$-$64 using the Las Cumbres Observatory (LCO) telescope network \citep{2013PASP..125.1031B}. The data were obtained using Johnson-Cousins $B$ ($3915 - 4806$~\AA) and $V$ ($5028 - 5868$~\AA) filters in an epoch between 11 April 2017 to 03 May 2017. Exposure times of 3\,s and 5\,s were used for the $V$ and $B$ filters, respectively. The raw images were reduced using the \textit{BANZAI} pipeline which, in summary, performs bad-pixel masking, bias subtraction, dark subtraction and flat-field correction \citep{2018SPIE10704E..01H}. Aperture photometry was used to extract the instrumental magnitudes using the \textsc{iraf} package \textit{apphot}. To calibrate the instrumental magnitudes to standard photometric systems, we used the known $B$ and $V$ magnitudes of the stars in the field of 4U 1210$-$64 from the \textit{APASS} catalogue \citep{2000A&A...355L..27H,2016yCat.2336....0H} and used the transformation equations described in \citet{1992ASPC...23...90D}. 

\subsubsection{Universitatsternwarte Bochum}
4U 1210$-$64 was observed using the Sloan $i$ ($6900 - 8400$~\AA) and $r$ ($5500 - 6900$~\AA) filters as part of the Bochum Survey of the Southern Galactic Disk catalog (\citealt{2015AN....336..590H}; labelled GDS 1213148$-$645230) between 12 May 2011 and 30 April 2015. The survey utilizes a robotic 15-cm twin telescope of the Universitatsternwarte Bochum in Chile. Several standard star fields were obtained each night and were used to perform the absolute calibration. The full details of the survey are described in \citet{2012AN....333..706H}.

\subsubsection{\textit{Swift}/UVOT}
We have also looked for available archival observations from the Ultra-Violet/Optical Telescope (UVOT, \citealt{Roming2005}) on board the \textit{Neil Gehrels Swift Observatory} \citep[hereafter \textit{Swift};][]{Gehrels2004} instrument. 
In this work, we have used the photometric observations in the three UW filters from the Swift Archive Download Portal.
 
\section{Results}
\begin{figure*}
\resizebox{\hsize}{!}{\includegraphics[width=15cm]{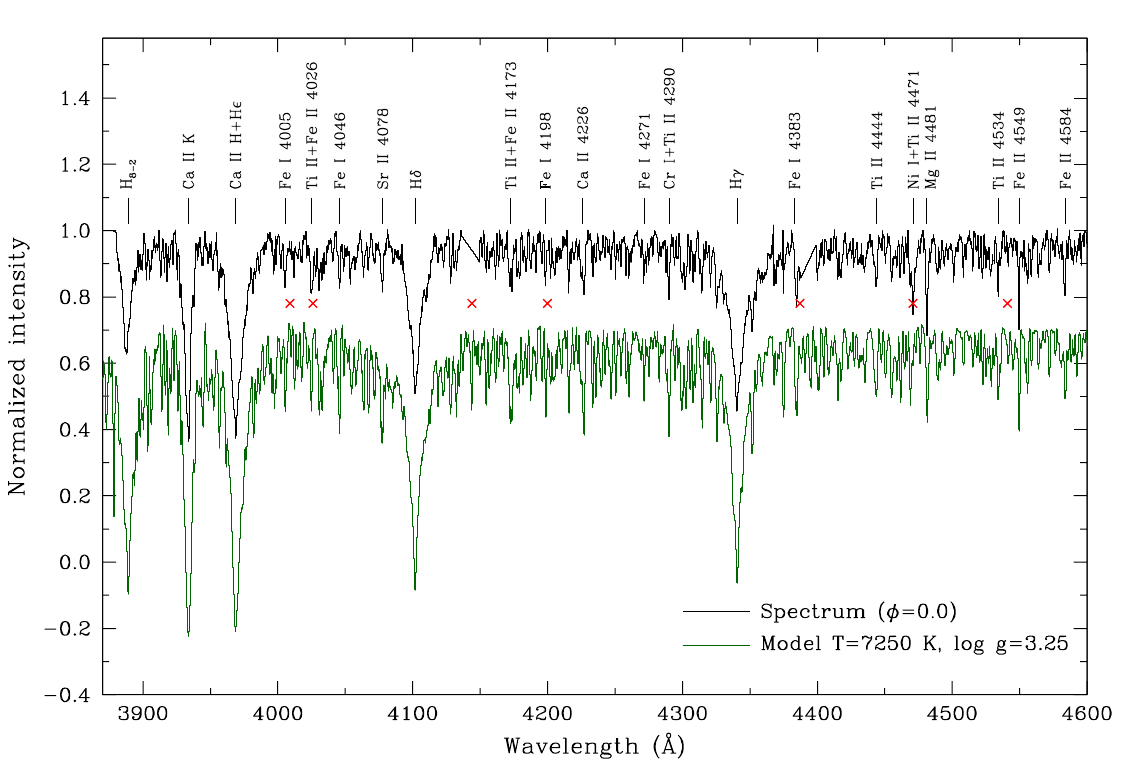}}
\caption{The normalized spectrum of 4U 1210-64 corresponding to
orbital phase zero (black), obtained on 12 Feb 2023 (MJD59988.02), and the 
Kurucz synthetic model computed with $T\!=\!7250$ K, $\log g\!=\!3.25$, solar
metallicity and $v \sin i\!=\!50$ km/s (green), also normalized, shifted 
0.4 units downwards for clarity (green). Some line identifications are included;
the red crosses mark the position at which the He {\sc i} lines would appear if
the object was a B-type star. See text for details.}
\label{fig:spec_synthefit}
\end{figure*}

\subsection{Characterization of the donor star}
\label{sec:donorstar}

In this section we carry out a thorough analysis of the spectra and  the spectral energy distribution (SED) of the system in order to quantify the parameters and the spectral classification of the donor star. For doing that, observations of the system, both spectroscopic and photometric, containing the minimum possible contribution of the accretion disk, have been used.
From the spectroscopy, we used the mid-resolution spectrum (PG3000) close to phase zero --when the star is eclipsing the compact object. The observation was taken on 12 February 2023 (MJD 59988.02), corresponding to an orbital phase 0.95.  The deep Ca~{\sc ii} K, and its ratio with the Ca~{\sc ii} H + H$\epsilon$ blend, together with the absence of He {\sc I} absorption lines, and the large number of metallic lines, suggest a spectral type around A8--9 \citep{Evans2003,GrayCorbally2009}, much later than the classification as a B-type star by \citet{Masetti2009}. The normalized spectrum is shown in Fig.\,\ref{fig:spec_synthefit} (lower panel) together with some line identifications; the red crosses mark the position where the He {\sc i} lines would appear, some of them very prominent, should the object be a B-type star. Note that the wavelengths of some of the strong absorptions in the spectrum match those of the He {\sc i} lines, however, according to the results of the spectral synthesis, the depths of the helium features are negligible as compared with those of the metals that appear in the labels.

Due to the complex characteristics of the system, we have obtained the fundamental parameters of the donor star through an iterative procedure in which the observed spectrum and the spectral energy distribution (SED), built with photometric data, are fitted in parallel; the combination of the results of each fit provides feedback for next step of the iteration until a self-consistent solution is reached for both sets of data.

The solution to the problem for this particular star is more complex than the case of a single object, since, as we will see below, the contribution of the disk and the close environment of the compact object to the total output of energy from the system, at all bands, is not negligible. Moreover, this contribution varies with the orbital phase and the state of the X-ray activity of the source. 

\subsubsection{Spectral fitting}
\label{subsec:spec}
To quantify the spectral classification and the stellar parameters, we first synthesized a set of high-resolution theoretical spectra, using the suite of programmes {\sc synthe} \citep{kurucz1993}, which takes as input the list of the atomic and molecular transitions and the spectral range to be synthesized, together with the corresponding model atmosphere for a given temperature, gravity and metallicity, which contains, among other ingredients, the atomic fractions of the different elements. The spectral resolution of the spectra collected does not allow to determine accurately the value of $v \sin i$, therefore we considered a standard value of 50\,km/s, typical of late-A stars. Solar metallicity was assumed.

Figure ~\ref{fig:spec_synthefit} shows the observed spectrum (12 Feb 2023, MJD 59988.02). Although at this epoch the source was in an active X-ray state (see Fig.\,\ref{fig:X_ray_lc}), we selected this spectrum because it has the highest available resolution in our dataset.
We explored the parameter space with $T_{\rm eff}$ between 7000 and 8000 K and $\log g$ between 2.5 and 4.5, and concluded that a model with $T_{\rm
  eff}$=7250 K, $\log g$=3.25 provides a good fit. It is apparent that bluewards below $\sim\!4300$ \AA{} there 
is a gradual decrease in the depth of the Balmer lines, the metal lines, 
and the Ca {\sc ii} K line, when compared with those of the synthetic 
spectrum. This veiling is most probably caused by a non-stellar contribution
which, when added to the stellar spectrum, has the effect of decreasing both the relative depth and the equivalent widths of the absorption lines. This hypothesis is reinforced by the fact that the He {\sc ii}\,4686 \AA\, is observed in emission at all orbital phases (even at phase 0) during the X-ray active epoch (2022/2023 spectra). Unfortunately, that wavelength does not fall within the interval covered by the observation obtained on 12 Feb 2023, but two lower-resolution spectra taken on 24 Jan 2023 (phase 0.10) and 31 Mar 2023 (phase 0.91) clearly show that He {\sc ii}\,4686 \AA\ line in emission, which implies that it must originate in a region with hot material. Therefore a plausible scenario is that even when the compact object is eclipsed by the donor star during the X-ray active state, a small fraction of the disc, probably the hot-spot, with material at a temperature larger than $\sim\!30,000$ K, could still be visible, contributing to increasing the blue continuum.

\subsubsection{Spectral energy distribution (SED) fitting}
\label{sec:SEDfit}
\begin{figure}
   \hspace{-0.25cm}\includegraphics[width=0.5\textwidth]{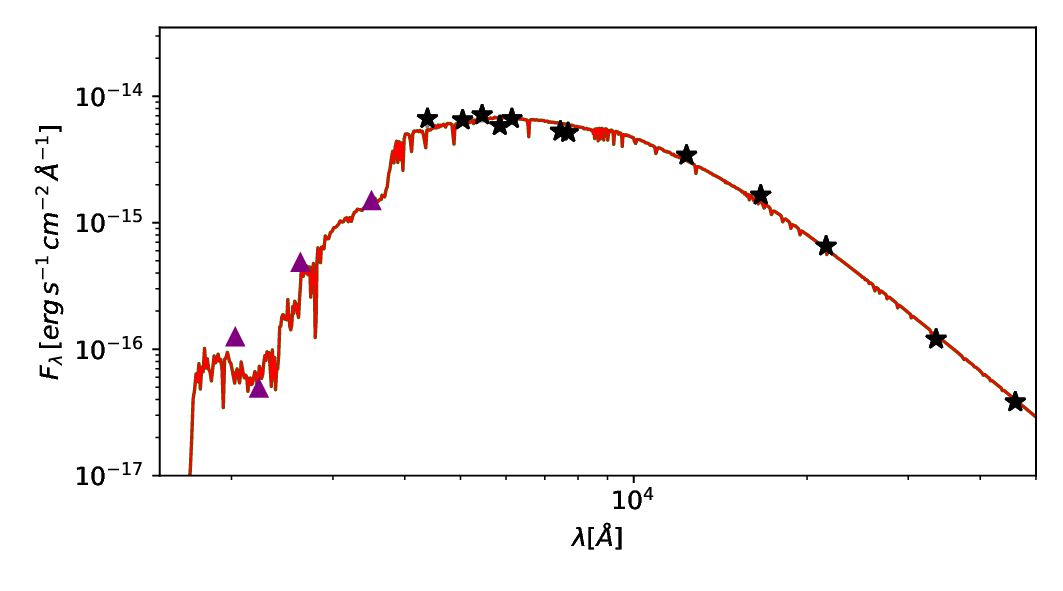}    
    \caption{Spectral Energy Distribution fit. Observations taken close to phase zero and in a low X-ray phase (see text), are plotted as black stars; the stellar theoretical flux reddened with the value which gives the best SED fit of $E(B\!-\!V)\!=\!0.60$ mag \textbf{is plotted in red}. The \textit{Swift}/UVOT UW2, UM2, UW1, and U mean fluxes from an archival observation from MJD~54580 are plotted as purple triangles for comparison.}
    \label{fig:Sed}
\end{figure}

\begin{figure}
\resizebox{\hsize}{!}{\includegraphics[width=15cm]{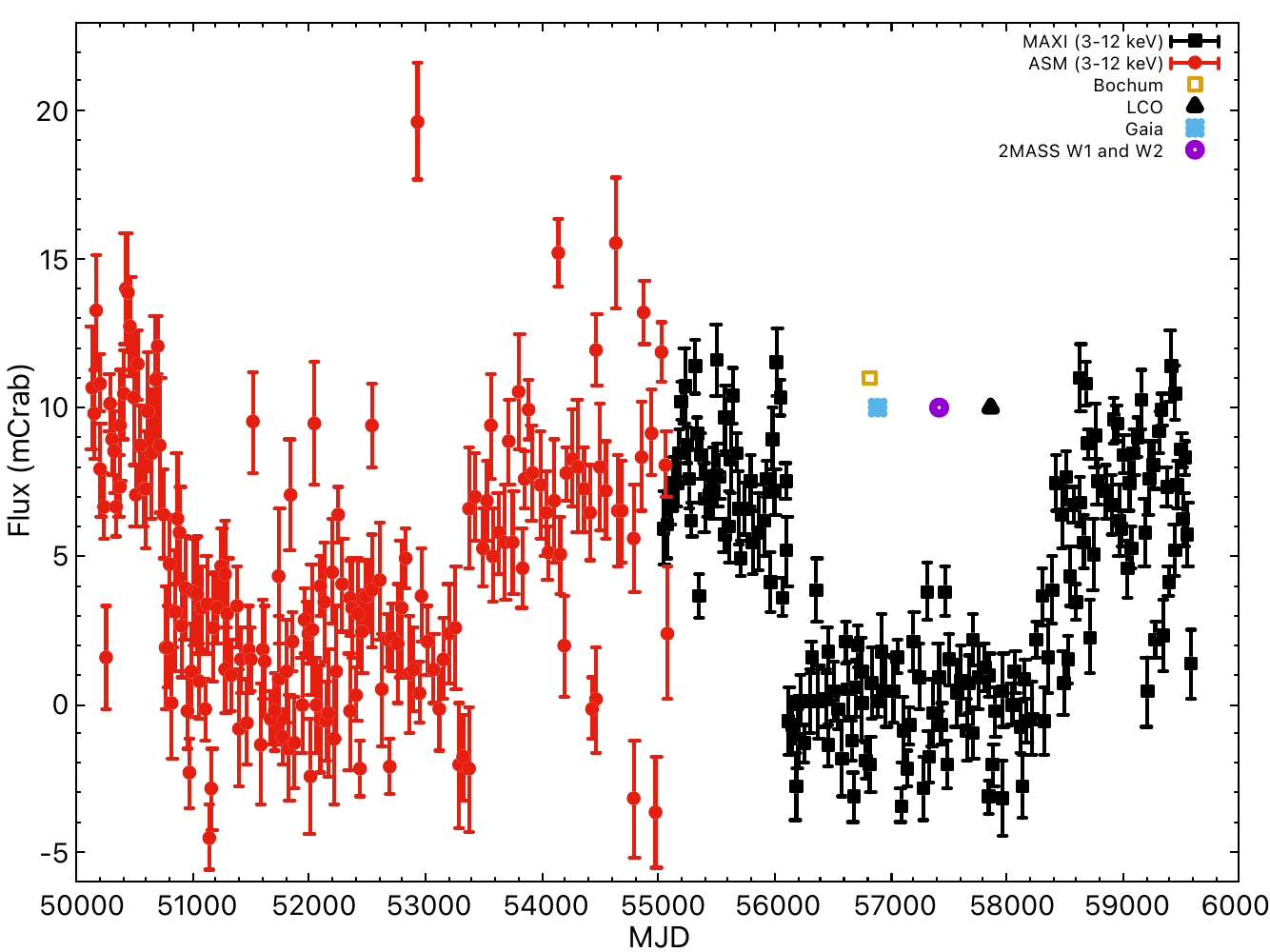}}
\caption{Long-term 3-12~keV \textit{RXTE} ASM and \textit{MAXI} lightcurve of 4U~1210$-$64. The lightcurve is binned to a time resolution of 21 days. The different symbols in the plot indicate the epochs of the photometric data that was used to perform the Spectral Energy Distribution fit.}
\label{fig:X_ray_lc}
\end{figure}

The spectral energy distribution (SED) has been built using photometric measurements obtained
during the least active states and/or close to zero orbital phase, when a minimum contribution from the emission of the accretion disk is expected (see Fig.~\ref{fig:Sed}). We selected the Gaia $BP$, $G$, and $RP$ observations at MJD 56887 (phase 0.80),  the Bochum $r$ and $i$ observations from MJD 56827.97 (phase 0.99), and the $B$ and $V$ LCO observations from MJD 57867.83 (phase 0.96). We also added the 2MASS $J$, $H$, and $K_{\rm s}$ measurements from the  2MASS All Sky Catalog of point sources \citep{Cutri2003} for which we do not know the dates of observations, although all the 2MASS observations were taken from June 1997 to February 2001, and the W1 and W2 measurements taken on MJD 57418.446 (phase 0.99) from the NEOWISE project \citep{Mainzer2011}.Additionaly, Swift/UVOT observations in the three UW and U filters obtained from MJD~54580.32 to 54580.99 are also plotted for reference, although since they were taken during an X-ray active phase, they were not used in the SED fit. The epochs of the photometric data used for the SED fit are indicated alongside the long-term X-ray lightcurve in Fig.~\ref{fig:X_ray_lc}.

We have used the distance from Gaia DR2 obtained by \citealt{Bailer-Jones2018}, who gave a value of $d\!=\!3.777$ kpc (16th percentile $d\!=\!3.530$ kpc, 84th percentile $d\!=\!4.064$ kpc); this value was preferred instead of that from Gaia eDR3 \citep{Bailer-Jones2021}, because it has a slightly smaller value of the RUWE (Renormalised Unit Weight Error) associated to each Gaia source \citep{Brown2021} (1.017 in DR2, against 1.046 in eDR3). The best SED
fit is obtained for R$_*$=6.4$\pm$0.2 R$_\odot$ and
$E(B\!-\!V)$=0.60$\pm$0.05\,mag (see Fig.\,\ref{fig:Sed}). This estimation of $E(B\!-\!V)$ is in agreement with the value
$E(B\!-\!V)$=0.66$\pm$0.06\,mag, obtained introducing the coordinates
and distance in the 3D interstellar dust reddening maps of the
Galactic plane by \citealt{Chen2019}.

Fig.~\ref{fig:Sed} shows the final fit to the SED. The photometric
observations are plotted as black stars, the stellar model multiplied by 4$\pi$R$_*$ (being R$_*$=6.4 R$_\odot$ the value providing the best fit), divided by 4$\pi$R$_{d}$ (being $d$=3.77\,kpc the distance from Gaia DR2), and reddened with the value of E(B-V) that provided the best fit ($E(B\!-\!V)$=0.60\,mag) is plotted in red.

\subsubsection{Stellar mass and final parameters}
\label{sec:mass_and_params}
\begin{figure}
    \centering
    \includegraphics[width=0.45\textwidth]{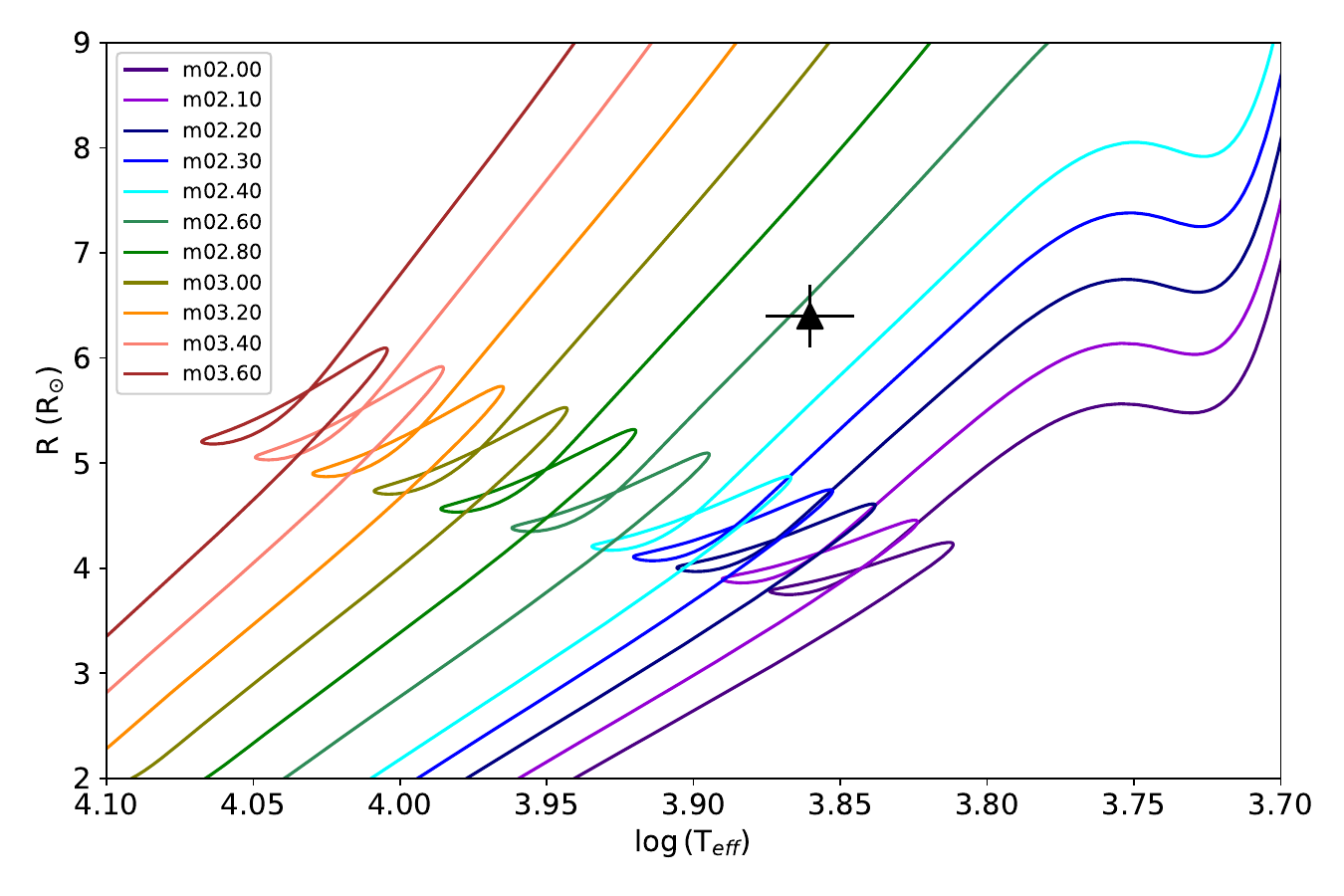}\\
    \includegraphics[width=0.45\textwidth]{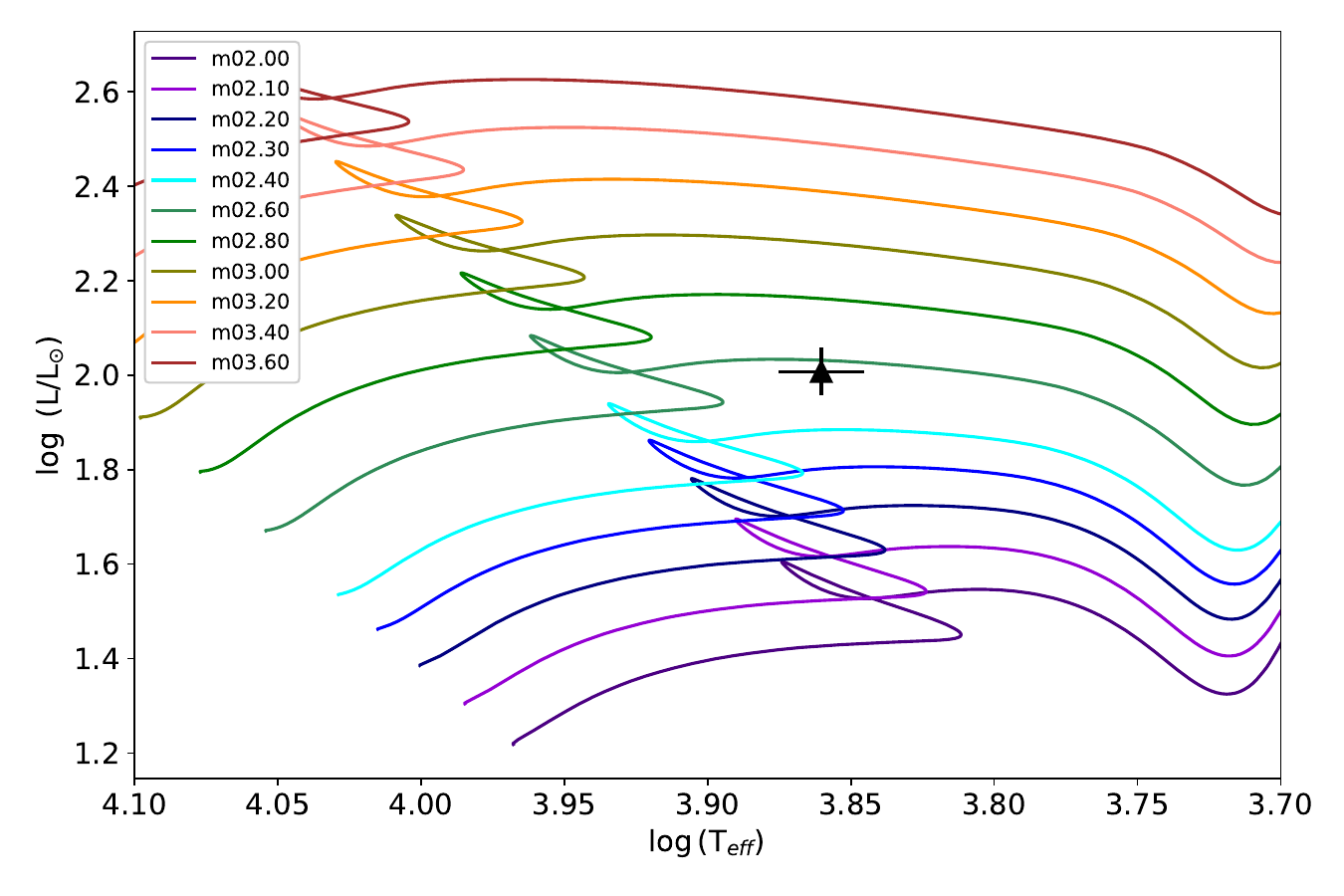}
    \caption{Evolutionary tracks for different masses in a
        R$_*$ vs $\log\,T_{\rm eff}$ diagram (top) and a 
        $\log\,(L/L_\odot$) vs $\log\,T_{\rm eff}$ (bottom) with the
        estimations from the SED fit marked with black triangles.}
    \label{fig:tracks}
\end{figure}

 The pair ($T_{\rm eff}$, R$_*$) --or, which is equivalent, ($T_{\rm eff}$, 
 L$_*$/L$_\odot$)-- was then superimposed on the set of PAdova and tRieste
 Stellar Evolution Code (PARSEC; \citealt{Bressan2012}) tracks and
 isochrones. Fig.~\ref{fig:tracks} shows the final position of the star in the HR diagrams, with the black triangle displaying the results from the best SED fit. The location of the star using the stellar parameters obtained in Sect.\,\ref{subsec:spec}, leads to an estimation of the mass M$_*$=2.55$\pm$0.05\,M$_\odot$. Using the radius 
obtained from the SED fitting, the gravity implied for this mass is $\log g
\!\simeq\!3.25$, i.e., which is in agreement with the spectroscopic estimation. We conclude that the final set of parameters for the donor star, with their corresponding conservative uncertainties is $T_{\rm eff}\!=\!7250\pm100$ K and $\log g\!=\!3.25\pm0.25$. Incidentally, from the orbital period of the system and the estimated value of the radius, under the assumption of synchronous rotation, the value of the equatorial stellar velocity would be $\sim$50\,km/s, which is in agreement with what we considered from the very beginning. 

Despite of the fact that both the spectral and SED fittings provide reasonable results, we are conscious that a deeper analysis,
mainly focused on the modeling of the disk and the environment
of the compact object, is required. That work, as well as a detailed study of the spectra and SED at different orbital phases is deferred to a follow-up paper. 
In conclusion, the spectral type of the optical companion can be
classified as $\sim$A8, and the location in the HR
diagram, implies a luminosity class around IV, i.e. the star seems to have left the main sequence, but not entered into the giant branch yet.

\subsection{Radial velocity analysis}
\label{sec:rv}

The relative radial velocities between the spectra of 4U~1210$-$64 were determined through cross-correlation analysis \citep{Tonry1979}. The method used is described by \citet{Foellmi2003}, \citet{Manick2015} and \citet{2017ApJ...847...68M} which uses the \textsc{xcsao} task in \textsc{iraf}. Heliocentric corrections were applied to the spectra using \textsc{iraf} tasks \textsc{rvcorr} and \textsc{dopcor}.

\begin{figure}
    \centering
    \includegraphics[width=0.4\textwidth]{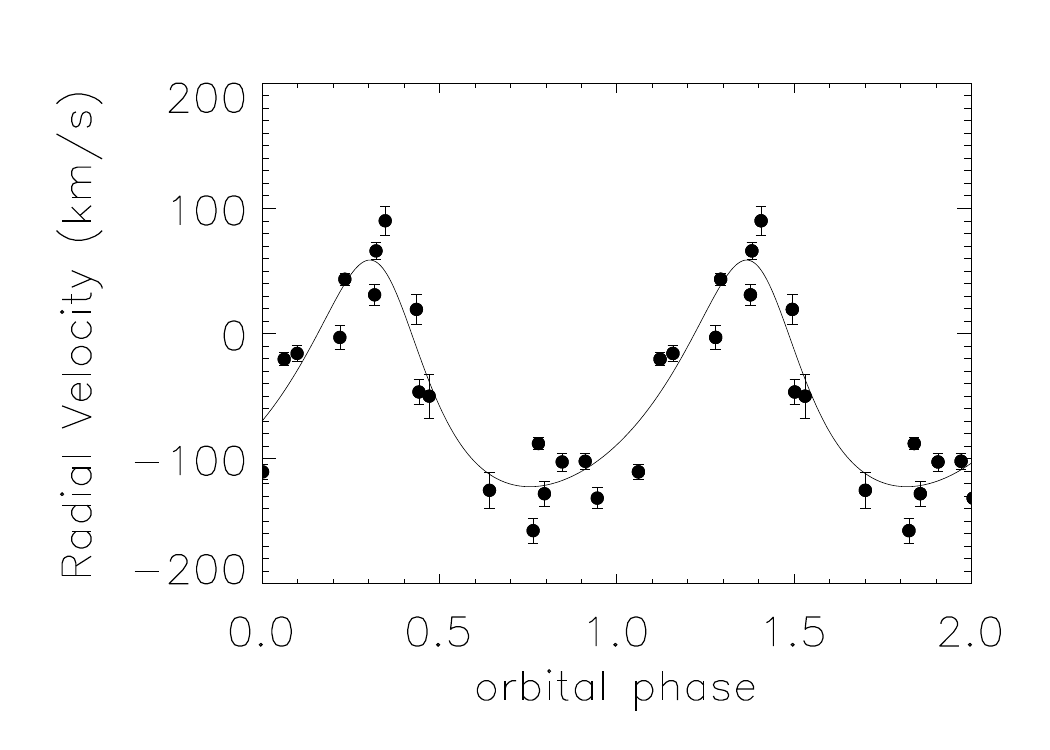}
   \includegraphics[height=2in]{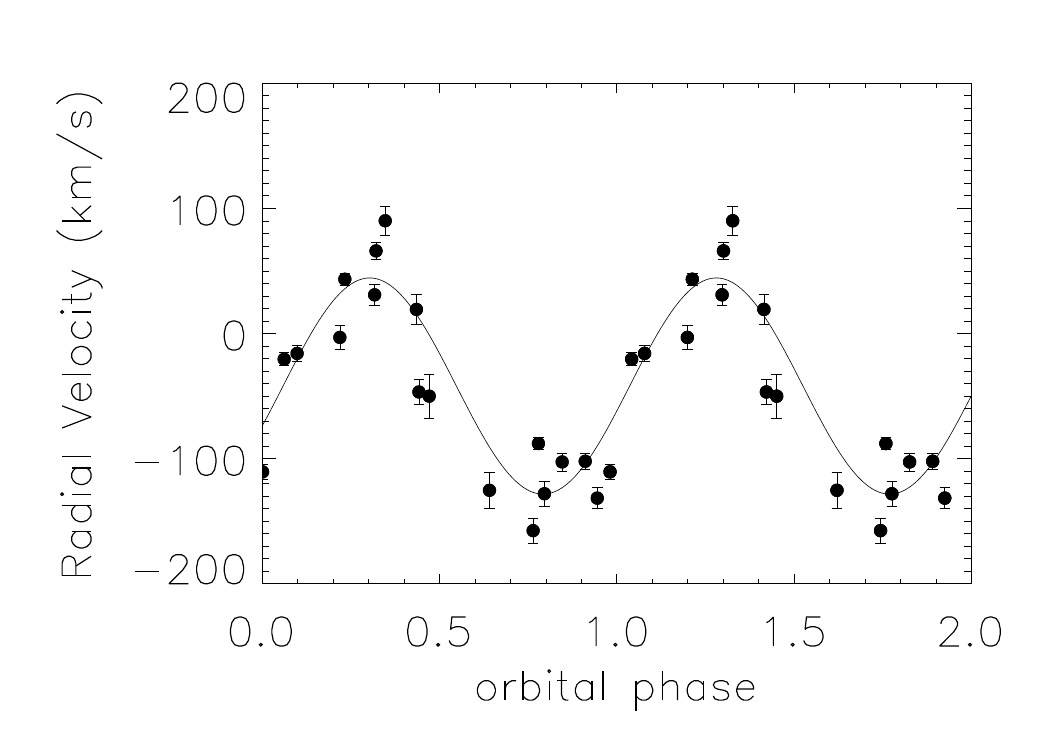}
    \caption{The measured radial velocity values folded on the orbital period with a Keplerian fit overplotted. The radial velocities plotted here are presented in Table~\ref{tab:RV_table}. Top panel, Case 1: a fixed orbital period and free eccentricity. Bottom panel, Case 2: a fixed orbital period but fixed eccentricity (circular orbit).}
    \label{fig:rv}
\end{figure}

We used an iterative process to generate a high signal-noise ratio (SNR) zero velocity template.  We firstly rectified and subtracted the continuum level of the spectra, after which they were converted to a logarithmic wavelength scale. The spectra were then grouped in descending order according to their SNR. We then measured the relative radial velocity shifts of each spectrum to the highest SNR spectrum. From the results of the first cross-correlation iteration, we shifted the individual spectra to the same wavelength as the high SNR template. The shifted spectra were then co-added, creating a mean spectrum which was used as a final template to compute the RV shifts shown in Table\,\ref{tab:RV_table} and Fig.\,\ref{fig:rv}.
\begin{table}
	\centering
	\caption{SALT observations of the optical counterpart of \hbox{4U 1210$-$64} with the measured radial velocities.}
	\label{tab:RV_table}
    \setlength\tabcolsep{2pt}
	\begin{tabular}{ccccc} 
		\hline\hline
		Date & JD & Orbital phase  & RV (km/s) & SNR \\
		\hline
2013-01-24 & 2456317.515 & 0.846 &  -103.0 $\pm$  7.5 & 49 \\
2013-02-06 & 2456330.602 & 0.796 &  -127.5 $\pm$  9.3 & 42\\
2013-02-26 & 2456350.614 & 0.779 &  -87.4 $\pm$  4.5 & 59 \\
2013-03-22 & 2456374.356 & 0.317 &  31.3 $\pm$  9.0 & 43\\
2013-04-12 & 2456395.324 & 0.442 &  -47.5 $\pm$  10.6 & 38 \\
2013-04-24 & 2456407.253 & 0.219 &  -2.6 $\pm$  9.4 & 62 \\
2014-01-18 & 2456676.515 & 0.347 &  90.3 $\pm$ 12.5 & 26 \\
2014-02-10 & 2456699.441 & 0.764 &  -156.5 $\pm$  9.9 & 41\\
2022-06-07 & 2459738.294 & 0.641 &  -125.2 $\pm$ 14.5 & 71\\
2022-06-19 & 2459750.328 & 0.435 &  19.1 $\pm$ 12.2 & 61 \\
2022-06-25 & 2459756.273 & 0.321 &  66.0 $\pm$ 6.7 & 82 \\
2022-06-26 & 2459757.282 & 0.471 &  -50.0 $\pm$ 17.9 & 94 \\
2022-06-29 & 2459760.233 & 0.911 &  -102.1 $\pm$ 6.4 & 62 \\
2022-06-30 & 2459761.246 & 0.062 &  -20.8 $\pm$ 5.5 & 53 \\
2022-07-13 & 2459774.260 & 0.001 &  -110.5 $\pm$ 5.1 & 54  \\
2022-07-28 & 2459789.232 & 0.233 &  44.2 $\pm$ 4.2 & 77 \\
2023-01-24 & 2459969.504 & 0.098 &  -16.0 $\pm$ 6.2 & 74 \\
2023-02-11 & 2459987.460 & 0.774 &  -210.6 $\pm$ 10.3 & 99 \\
2023-02-12 & 2459988.602 & 0.945 &  -131.6 $\pm$ 8.3 & 23 \\
		\hline
	\end{tabular}
\end{table}

\begin{table}
	\centering
	\caption{Orbital parameters derived from Keplerian fits to the radial velocities for the two cases studied.}
	\label{tab:orb_pars1}
	\begin{tabular}{l  c  c} 
		\hline\hline
		Parameter & Value  & Value  \\
		 & \textbf{Case 1} &  \textbf{Case 2} \\
		\hline
		Period (days) & 6.7101 (fixed) & 6.7101 (fixed)  \\
		Eccentricity & 0.27 $\pm$ 0.23 & 0.00 (fixed) \\
		$\omega$ (degrees) & 25 $\pm$ 13 & 51 $\pm$ 12 \\
		K (km/s) & 91 $\pm$ 20 & 86 $\pm$ 15 \\
		$\gamma$ (km/s) & -54 $\pm$ 12  & -42 $\pm$ 11\\
		\hline
	\end{tabular}
\end{table}

Due to the modest orbital phase coverage we performed Keplerian fits to the radial velocity measurements with different settings. In Case 1 we fixed the period to that obtained from X-ray analysis \citep{Coley2014} but with the eccentricity left as a free parameter. In Case 2, both the period and eccentricity were fixed to perform a sinusoidal fit. Fig.\,\ref{fig:rv} shows the fits for the two cases, with the orbital parameters listed in Table\,\ref{tab:orb_pars1}, where $\omega, K$ and $\gamma$ are the longitude of periastron, semi-amplitude and systemic velocity of the binary, respectively. In our notation, the longitude of periastron is the sum of the argument of periastron and the longitude of the ascending node. The mass function can then be calculated using

\begin{equation}
\label{MF_eqn}
f(M) = \frac{M_{1}^{3}}{(M_{1}+M_{2})^{2}} \sin^{3}i = \frac{P}{2\pi G}K_{2}^{3}(1-e^2)^{3/2}
\end{equation}
where:
\begin{itemize}
\item $M_1 = $ compact object mass
\item $M_2 = $ donor star mass
\item $K_2 = $ semi-amplitude velocity of the donor star
\item $P = 6.7101 $d \citep{Coley2014}
\end{itemize}
For the parameters listed in Table~\ref{tab:orb_pars1}, a mass function of $f(M) = 0.47 \pm 0.32$~M$_\odot$ and $f(M) = 0.44 \pm 0.22$~M$_\odot$ is obtained for the case of the eccentricity left as a free parameter (Case 1) and the eccentricity fixed (Case 2), respectively.

\subsection{Folded light curves}
\label{sec:photometry}
\begin{figure*}
\includegraphics[width=\textwidth,height=0.95\textheight,angle=0]{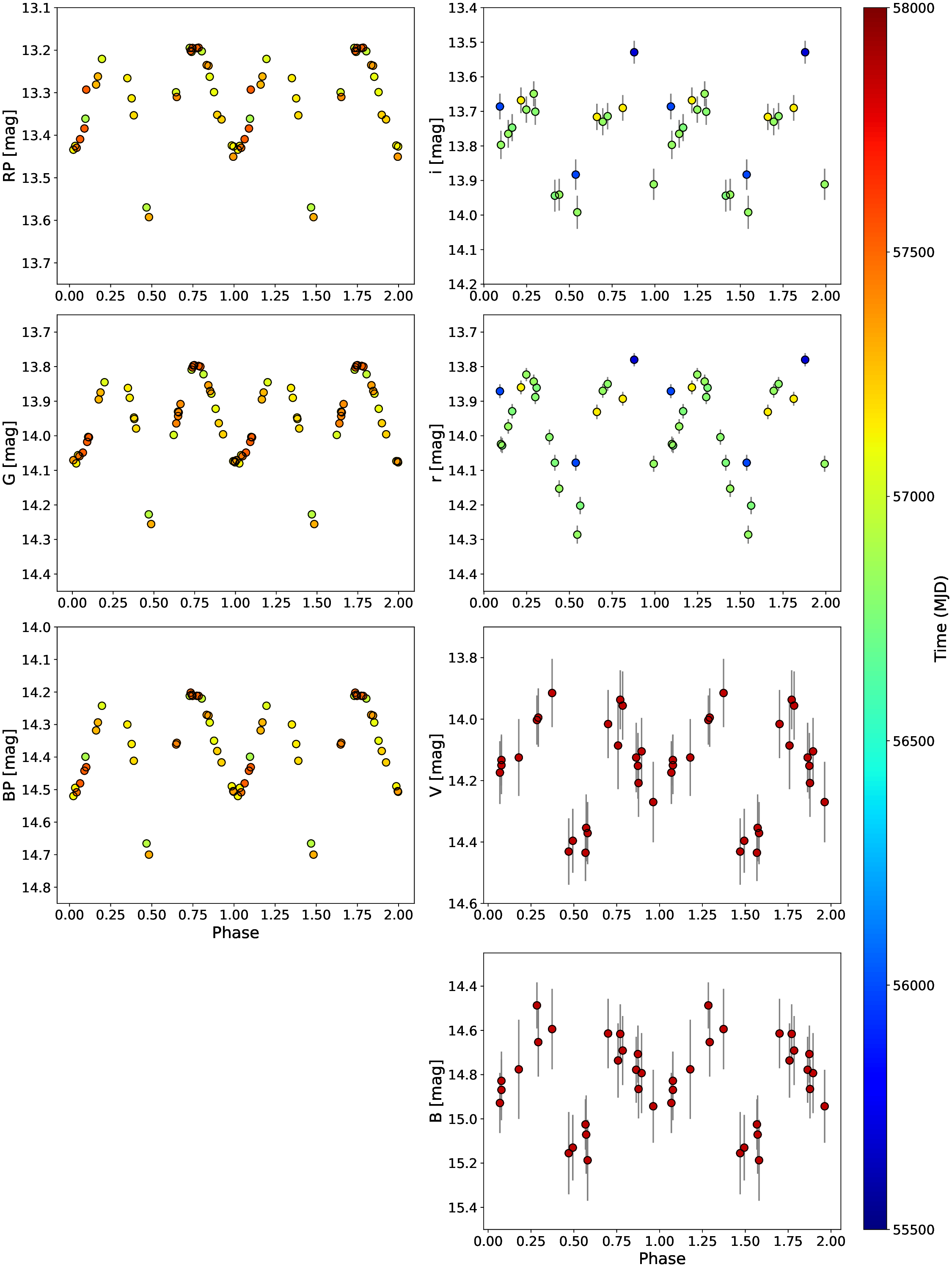}
\caption{Folded optical light curves. On the left, from top to bottom: Gaia RP, Gaia G, Gaia BP; on the right, from top to bottom: Bochum r, Bochum i, LCO V, LCO B. The folded light-curves are colour-coded with time (see colorbar). Blue points are from an X-ray active phase (see \citealt{Coley2014}).}
\label{fig:folded_LCs}
\end{figure*}

Fig.~\ref{fig:folded_LCs} shows the optical light curves from the observations described in Sect.\ref{subsec:phot} folded over the orbital period using the ephemeris described in \citet{Coley2014}.
Although the various optical datasets were taken at different epochs, the modulation of the lightcurves shows similar behaviour. The folded lightcurves show sinusoidal modulation with a dip in flux between orbital phases $\phi \sim 0.4$ and $\phi \sim 0.6$, similar to that observed in the X-ray lightcurve between phases $\phi \sim - 0.04$ and $\phi \sim 0.03$ \citep{Coley2014}. Analogous to the X-rays, the dip in flux seen in the optical lightcurves is suggestive of an eclipse. The shape of the optical lightcurves is possibly due to a combination of ellipsoidal variations, irradiation effects and eclipsing of the two stellar components of the binary. The minimum at $\phi \sim 0.5$ is noticeably deeper than that at $\phi \sim 0$. At $\phi \sim 0.5$ the accretion disc eclipses the hotter, irradiated side of the donor star while at $\phi \sim 0$ the observer views the cooler, non-irradiated side of the star when the compact object and accretion disc are eclipsed. In Fig.~\ref{fig:sketch} an illustration of the binary system of 4U 1210$-$64 is provided.

\section{Discussion}
\label{sec:discussion}
\subsection{An intermediate-mass X-ray binary}
The firm classification of the optical counterpart of \hbox{4U 1210$-$64} as an A8~III-IV star means that this object is a rare intermediate-mass X-ray binary (IMXB). These systems are rare because their mass transfer timescales are short. As explained by \citet{Tauris2006}, the significantly higher mass ratio in IMXBs means that when Roche lobe overflow does occur in the system, it happens rapidly and is either over on a sub-thermal timescale or only in a 1000 years. The weak stellar wind outflow from the donor star is also insufficient to power the compact object for strong X-ray emission to be detected. This presents a selection bias against detecting IMXBs as X-ray sources.

\begin{figure*}
   \includegraphics[width=0.8\textwidth]{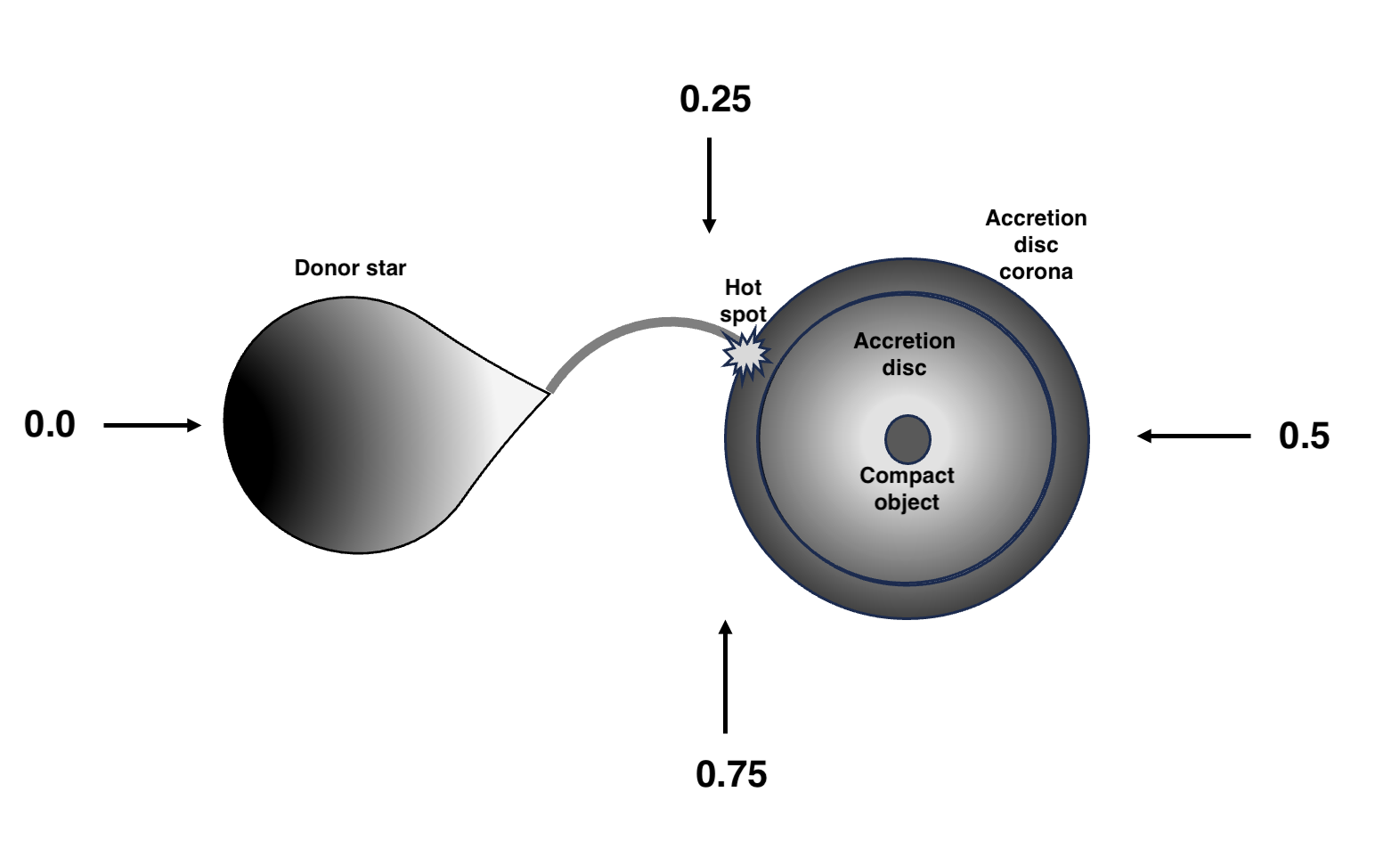}  
    \caption{A schematic illustrating a face-on view of the binary system with the components labeled. The binary phases are indicated.}
    \label{fig:sketch}
\end{figure*}

\subsection{Accretion disc corona}
The X-ray lightcurve of \hbox{4U 1210$-$64} shows a smooth modulation with a minimum occurring $\sim 0.2$ cycles before the eclipse \citep{Coley2014}. Similar modulation has been observed in the low-mass X-ray binaries, 4U~1822$-$37 and 4U~2129$+$47 \citep{1982ApJ...257..318W,2000A&A...356..175P}. The eclipse can be interpreted as an X-ray emission region that is extended due to the X-rays being scattered in the accretion disc corona (ADC).
The smooth variation is caused by X-rays from ADC being obscured by the structure at the rim of the accretion disc. According to \citet{Coley2014}, the X-ray lightcurve shows an eclipse that spans an orbital phase range between $\phi \sim -0.04$ and $\phi \sim 0.03$. The optical photometric lightcurve shows an eclipse almost half an orbital phase later spanning orbital phase range $\phi \sim 0.4$ and $\phi \sim 0.6$. During the X-ray eclipse the donor star completely obscures the compact object and the extended corona as it is projected onto our line of sight (phase 0.0 in Fig.~\ref{fig:sketch}). The X-ray eclipse has a shorter time span than the optical eclipse. This is possibly due to the hot spot, which is responsible for significant X-ray emission, being compact. In this picture, when the donor star eclipses the compact object and accretion disc during the X-ray eclipse it blocks out the hot spot briefly while the accretion disc blocks out a large portion of the optical light from the donor star for a longer period.\\

\subsection{Nature of the compact object}
\begin{figure}
    \centering
    \includegraphics[width=0.4\textwidth]{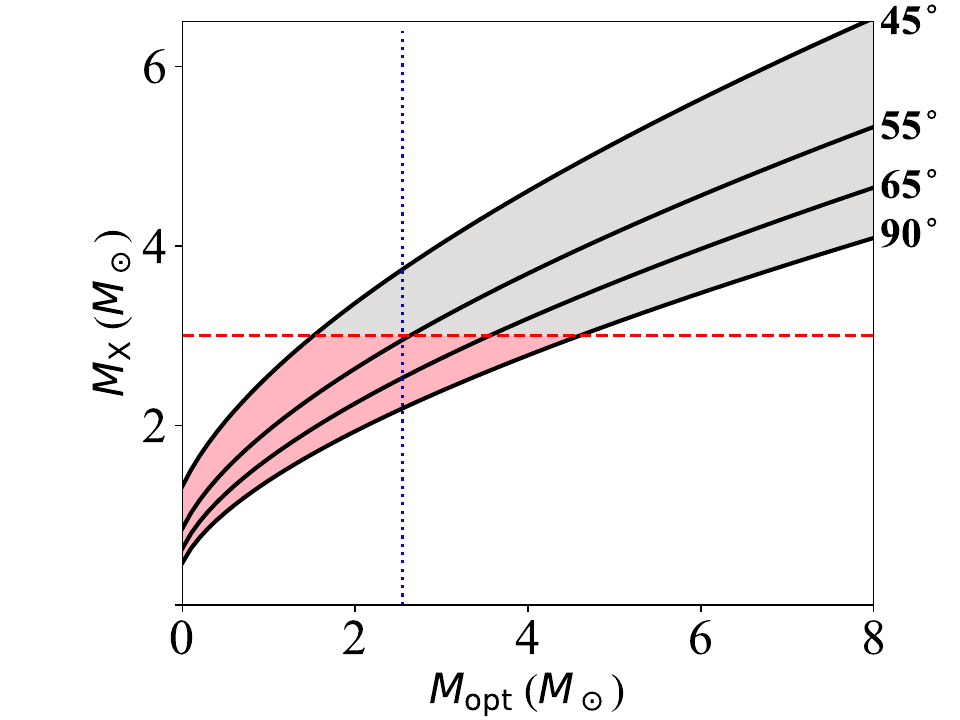}\\
    \includegraphics[width=0.4\textwidth]{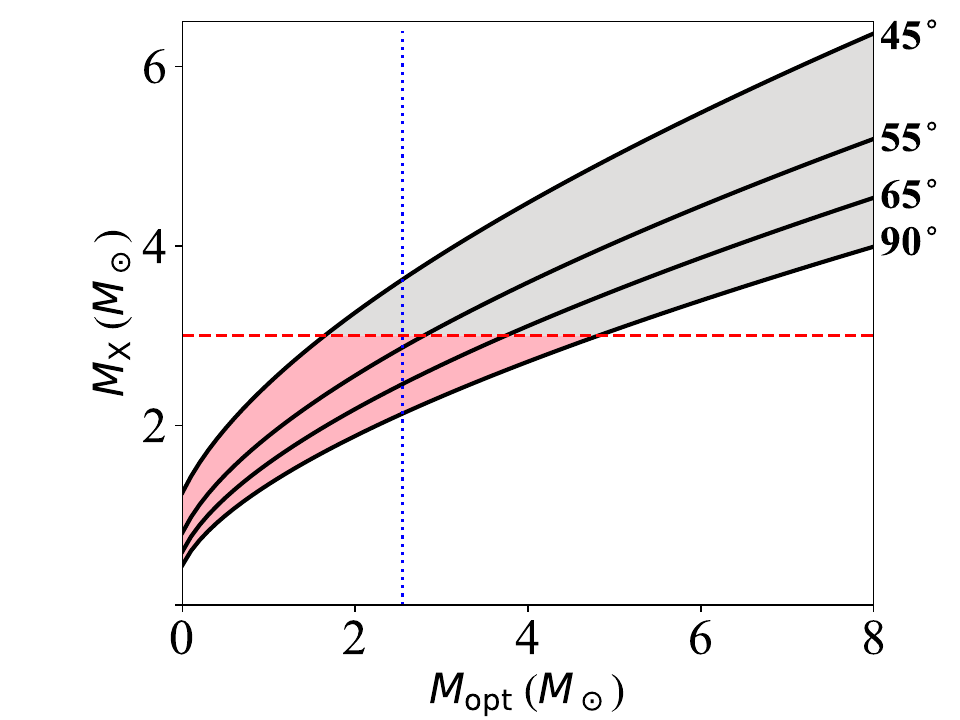}
    \caption{Mass-mass plot from the calculated mass-function showing the mass constraints of the two objects as a function of inclination angle for the sinusoidal Keplerian fit to the radial velocity measurements. The vertical line in both plots indicates the mass of the donor star of 2.55 M$_\odot$. Top panel, Case 1: a fixed orbital period and free eccentricity. Bottom panel, Case 2: a fixed orbital period but fixed eccentricity (circular orbit).}
    \label{fig:mass_x_mass_opt}
\end{figure}

Fig.\,\ref{fig:mass_x_mass_opt} shows the mass-mass plots obtained from the orbital parameters listed in Table~\ref{tab:orb_pars1} and equation\,\ref{MF_eqn}, with the mass constraints of the two stars for different inclination angles for the calculated mass function. The results are similar for both cases of the eccentricity left as a free parameter (Case 1) and the sinusoidal fit (Case 2). Using the donor star mass obtained in Section~\ref{sec:mass_and_params} of 2.55~M$_\odot$ and a range of compact object masses of 1.4~M$_\odot$ (canonical neutron star mass) and 2.35~M$_\odot$ (the most massive neutron star known; \cite{Romani2022}), we find that a neutron star is possible for inclination angles in the range $\sim 50^\circ \lesssim i \lesssim 73^\circ$. For the same donor star mass, a black hole ($M_X \gtrsim 3.0$~M$_\odot$) is possible for an inclination angle range $i \lesssim 54^\circ$. The eclipses that are seen in the optical and X-ray lightcurves suggest that the plane of the orbit is viewed at large inclination angles close to $90^\circ$, which favours a neutron star as the compact object in the binary system. It is also possible to use the mass functions obtained for Case 1 and Case 2 with an inclination angle of $i = 90^\circ$ to estimate the mass range of the donor star. For Case 1, a lower limit canonical neutron star mass of 1.4~M$_\odot$ results in a donor star mass of $M_\ast = 1.0$~M$_\odot$, while the upper limit neutron star mass of 2.35~M$_\odot$ yields $M_\ast = 2.9$~M$_\odot$. Similarly for Case 2 the lower and upper limits of the neutron star mass result in donor star masses of $M_\ast = 1.1$~M$_\odot$ and $M_\ast = 3.1$~M$_\odot$, respectively. A black hole ($M_X \gtrsim 3.0$~M$_\odot$) results in donor star mass range $M_\ast \gtrsim 4.6$~M$_\odot$.

Binary stellar evolution models indicate that donor stars in binary systems exhibit different properties to isolated stars of the same spectral type as a result of mass transfer to the companion \citep{Rappaport1983}. This results in undermassive and undersized donor stars for a given spectral type. The mass of the donor star obtained in Section~\ref{sec:mass_and_params} uses isochrone fitting models that do not take into account binary evolution. Consequently, the mass estimate obtained from isochrone fitting could be lower, which would make the neutron star even more probable.

\section{Conclusions}
\label{sec:conclusion}
We have presented optical spectroscopic and photometric observations of 4U 1210$-$64 obtained with SALT and LCO/Bochum/Gaia, respectively. From the analysis of the spectra, we demonstrate that the donor star is an A8~III-IV spectral type, making the system a member of the rare intermediate-mass binary sub-class. The shape of the phase-folded optical photometric lightcurves reveals behaviour that is possibly due to ellipsoidal variations, irradiation of the donor star, and mutual eclipses of the star and accretion disk. We have measured radial velocities from the spectra and fitted a Keplerian model where we have considered eccentric and sinusoidal fits. Based on the results of the fits, we suggest that a neutron star is strongly favoured as the compact object.

\section*{Acknowledgements}

LJT is supported by the SALT Foundation and the South African NRF. IMM and VAM are supported by the South African NRF. JAG is funded by Spanish
MCIN/AEI/10.13039/501100011033 grant PID2019-107061GB-C61. BM is supported by grant
PID2021-127289-NB-I00 by the Spanish Ministry of Science and Innovation
(MCIN) / State Agency of Research (AEI). The material is based upon work supported by NASA under award number 80GSFC21M0002. Some of the observations reported in this paper were obtained with the Southern African Large Telescope (SALT), under the science programmes: 2012-2-RSA\_OTH\_UKSC-003, 2013-2-RSA\_OTH\_UKSC-003 (PI: McBride) and 2021-2-MLT-005 (PI: Townsend).

\section*{Data Availability}

The X-ray data are freely available from the RXTE and MAXI archives. The SALT optical data in this article will be shared on any reasonable request to Vanessa McBride and Lee Townsend. The LCO data will be shared on any reasonable request to Itumeleng Monageng. The Gaia and Bochum data are freely available from the missions' respective webpages.



\bibliographystyle{mnras}
\bibliography{example} 




\appendix

\section{Dates for X-ray and optical observations}
\begin{table}
	\centering
	\caption{A log of the Bochum photometric observations.}
	\label{tab:Bochum}
    \setlength\tabcolsep{2pt}
	\begin{tabular}{cccc} 
		\hline\hline
		Date (MJD) & Orbital Phase & i band mag  & r band mag \\
		\hline
55693.1991200 &  0.8791404 &   13.780 $\pm$  0.019	&	13.529 $\pm$   0.033 \\		  
55983.1720000 &  0.0935306 &   13.871 $\pm$  0.020 &	13.686 $\pm$   0.037 \\	  		
55986.1487000 &  0.5371455 &   14.078 $\pm$  0.023 &	13.883 $\pm$   0.044 \\  		
56721.3698000 &  0.1064664 &   14.028 $\pm$  0.022 &	  -		 \\
56750.2800800 &  0.4149387 &   14.078 $\pm$  0.023 &	13.944 $\pm$   0.046  \\		
56751.2750900 &  0.5632241 &   14.202 $\pm$  0.025 &	  -		\\
56755.3153100 &  0.1653343 &   13.929 $\pm$  0.021 &	13.747 $\pm$   0.039 \\  	
56756.2745300 &  0.3082860 &   13.861 $\pm$  0.020 &	  -	 \\
56783.0629500 &  0.3005395 &   13.888 $\pm$  0.020	&	13.701 $\pm$   0.038  \\ 	
56797.0383900 &  0.3832864 &   14.004 $\pm$  0.022 &	  -	\\
56806.0334500 &  0.7238119 &   13.850 $\pm$  0.020	&	13.714 $\pm$   0.038  \\ 	
56821.9745700 &  0.0995022 &   14.024 $\pm$  0.022 &	13.797 $\pm$   0.041 \\	  	
56822.9715500 &  0.2480813 &   13.823 $\pm$  0.019 &	13.695 $\pm$   0.038  \\ 	
56824.9718300 &  0.5461811 &   14.286 $\pm$  0.026 &	13.992 $\pm$   0.048 \\	  	
56825.9719800 &  0.6952326 &   13.870 $\pm$  0.020 &	13.730 $\pm$   0.039  \\ 	
56827.9720000 &  0.9932937 &   14.081 $\pm$  0.023	&	13.911 $\pm$   0.045  \\ 	
56828.9721200 &  0.1423407 &   13.973 $\pm$  0.021 &	13.765 $\pm$   0.040  \\ 	
56829.9723000 &  0.2913966 &   13.843 $\pm$  0.020 &	13.649 $\pm$   0.036  \\ 	
56830.9726700 &  0.4404808 &   14.153 $\pm$  0.024 &	13.941 $\pm$   0.046  \\ 	
57138.1357100 &  0.2167047 &   13.860 $\pm$  0.020 &	13.668 $\pm$   0.037  \\ 	
57141.1161700 &  0.6608799 &   13.931 $\pm$  0.021 &	13.716 $\pm$   0.038  \\ 	
57142.1317500 &  0.8122308 &   13.893 $\pm$  0.020 &	13.690 $\pm$   0.037  \\ 	
		\hline
	\end{tabular}
\end{table}
\begin{table}
	\centering
	\caption{A log of the LCO photometric observations.}
	\label{tab:LCO}
    \setlength\tabcolsep{2pt}
	\begin{tabular}{cccc} 
		\hline\hline
		Date (MJD) & Orbital Phase & V band mag  & B band mag \\
		\hline
57855.1830477 &  0.0776065 &    14.133 $\pm$   0.083 &      14.828 $\pm$   0.132  \\	
57858.5004219 &  0.5719918 &    14.354 $\pm$   0.109 &      15.071 $\pm$   0.177  \\	
57862.5733099 &  0.1789705 &    14.125 $\pm$   0.125 &      14.776 $\pm$   0.224  \\	
57860.5255085 &  0.8737885 &    14.152 $\pm$   0.107 &      14.707 $\pm$   0.129  \\	
57857.1669306 &  0.3732628 &    13.915 $\pm$   0.111 &      14.594 $\pm$   0.182  \\	
57859.3580015 &  0.6997961 &    14.016 $\pm$   0.111 &      14.614 $\pm$   0.157  \\	
57864.5277377 &  0.4702371 &    14.431 $\pm$   0.108 &      15.155 $\pm$   0.186  \\	
57866.6387006 &  0.7848319 &    13.956 $\pm$   0.111 &      14.691 $\pm$   0.156  \\	
57861.8337478 &  0.0687542 &    14.174 $\pm$   0.102 &      14.928 $\pm$   0.136  \\	
57853.9592288 &  0.8952219 &    14.105 $\pm$   0.109 &      14.793 $\pm$   0.181  \\	
57859.8337043 &  0.7706896 &    13.937 $\pm$   0.096 &      14.616 $\pm$   0.134  \\	
57867.8337022 &  0.9629219 &    14.270 $\pm$   0.131 &      14.943 $\pm$   0.165  \\	
57876.7023838 &  0.2846133 &    14.003 $\pm$   0.080 &      14.487 $\pm$   0.104  \\	
57873.1677888 &  0.7578559 &    14.086 $\pm$   0.142 &      14.736 $\pm$   0.168  \\	
57873.9694145 &  0.8773214 &    14.208 $\pm$   0.110 &      14.865 $\pm$   0.133  \\	
57875.9742466 &  0.1760997 &    14.447 $\pm$   0.131 &      15.085 $\pm$   0.182  \\	
57865.1830831 &  0.5679026 &    14.435 $\pm$   0.092 &      15.025 $\pm$   0.114  \\		
57867.1670566 &  0.8635723 &    14.125 $\pm$   0.113 &      14.778 $\pm$   0.149  \\		
57855.1837082 &  0.0777050 &    14.150 $\pm$   0.094 &      14.869 $\pm$   0.137  \\		
57870.0454548 &  0.2925373 &    13.995 $\pm$   0.095 &      14.653 $\pm$   0.156  \\		
57871.9758040 &  0.5802155 &    14.371 $\pm$   0.101 &      15.187 $\pm$   0.183  \\		
57871.3967696 &  0.4939225 &    14.396 $\pm$   0.104 &      15.130 $\pm$   0.148  \\		
		\hline
	\end{tabular}
\end{table}


\bsp	
\label{lastpage}
\end{document}